\documentclass[10pt,conference]{IEEEtran}

\usepackage{amssymb, stmaryrd}
\usepackage{amsfonts}
\usepackage{amsmath}
\usepackage{latexsym}
\usepackage{epsfig}
\usepackage{psfrag}
\usepackage{bm}
\usepackage{xspace}
\usepackage{graphicx}

\renewcommand{\baselinestretch}{0.995}

\newtheorem{Definition}{Definition}
\newtheorem{Lemma}[Definition]{Lemma}
\newtheorem{Theorem}[Definition]{Theorem}
\newtheorem{Example}[Definition]{Example}
\newtheorem{Assumption}[Definition]{Assumption}

\newtheorem{Corollary}[Definition]{Corollary}

\newenvironment{Proof}%
  {\noindent \emph{Proof:}}{\hfill$\square$}

\newcommand{\eexample}{\mbox{ }\hfill$\square$}
\newcommand{\eassumption}{\hfill$\square$}
\newcommand{\edefinition}{\hfill$\square$}

\newcommand{\ignore}[1]{}

\newcommand{\vomega}{\boldsymbol{\omega}}

\newcommand{\hvomega}{\hat{\boldsymbol{\omega}}}
\newcommand{\vGamma}{\boldsymbol{\Gamma}}
\newcommand{\vgamma}{\boldsymbol{\gamma}}

\newcommand{\matr}[1]{\mathbf{#1}}
\newcommand{\vect}[1]{\mathbf{#1}}

\newcommand{\vX}{\vect{X}}
\newcommand{\vx}{\vect{x}}

\newcommand{\vY}{\vect{Y}}
\newcommand{\vy}{\vect{y}}

\newcommand{\tr}{\mathsf{T}}

\newcommand{\set}[1]{\mathcal{#1}}
\newcommand{\code}[1]{\mathcal{#1}}
\newcommand{\defeq}{\triangleq}

\newcommand{\convhull}{\operatorname{conv}}
\newcommand{\conichull}{\operatorname{conic}}

\newcommand{\Gammapos}{\Gamma_{\mathrm{pos}}}
\newcommand{\gammapos}{\gamma_{\mathrm{pos}}}
\newcommand{\Gammaneg}{\Gamma_{\mathrm{neg}}}
\newcommand{\gammaneg}{\gamma_{\mathrm{neg}}}

\newcommand{\Expec}{\operatorname{E}}
\newcommand{\Prob}{\operatorname{Pr}}
\newcommand{\dint}[1]{\operatorname{d}\!#1}

\newcommand{\fp}[1]{\set{#1}}
\newcommand{\fph}[2]{\set{#1}(\matr{#2})}
\newcommand{\fc}[1]{\set{#1}}
\newcommand{\fch}[2]{\set{#1}(\matr{#2})}

\newcommand{\wcol}{w_{\mathrm{col}}}
\newcommand{\wrow}{w_{\mathrm{row}}}

\newcommand{\GF}[1]{\mathbb{F}_{#1}}
\newcommand{\R}{\mathbb{R}}
\newcommand{\Rp}{\mathbb{R}_{+}}
\newcommand{\Rpp}{\mathbb{R}_{++}}
\newcommand{\card}[1]{\lvert #1 \rvert}

\newcommand{\Ec}{E_{\mathrm{c}}}

\newcommand{\intextbinomial}[2]{\mbox{ \scriptsize $\!\!\left(\!\!
      \begin{array}{c}
        #1 \\
        #2
      \end{array} \!\!\right)\!$}}

\begin{document}

\title{Bounds on the Threshold of \\
       Linear Programming Decoding}

\author{
\authorblockN{Pascal O.~Vontobel}
\authorblockA{Dept.~of EECS\\
              Massachusetts Institute of Technology\\
              Cambridge, MA 02139, USA\\
              \texttt{pascal.vontobel@ieee.org}}
\and
\authorblockN{Ralf Koetter}
\authorblockA{CSL and Dept.~of ECE\\
              University of Illinois at Urbana-Champaign\\
              Urbana, IL 61801, USA\\
              \texttt{koetter@uiuc.edu}}
}

\maketitle

\begin{abstract}
  Whereas many results are known about thresholds for ensembles of low-density
  parity-check codes under message-passing iterative decoding, this is not the
  case for linear programming decoding. Towards closing this knowledge gap,
  this paper presents some bounds on the thresholds of low-density
  parity-check code ensembles under linear programming decoding.
\end{abstract}

\section{Introduction}
\label{sec:introduction:1}

Message-passing iterative (MPI) decoding and linear programming (LP) decoding
are very efficient methods to achieve excellent decoding performance of
low-density parity-check (LDPC) codes on a variety of channels. While an
enormous amount of work has been devoted to the understanding of LDPC codes
under MPI decoding (see e.g.~\cite{Wiberg:96,
Luby:Mitzenmacher:Shokrollahi:Spielman:Stemann:97:1,
Luby:Mitzenmacher:Shokrollahi:98:1, Richardson:Urbanke:01:2,
Lentmaier:Truhachev:Costello:Zigangirov:04:1, Jin:Richardson:05:1} for results
on thresholds), comparably few results on the performance of LDPC codes using
the more recent LP decoding are known. In this paper we provide analytical
bounds on thresholds of LP decoding; these bounds establish necessary
conditions for the existence of LP decoding thresholds.

We note that the existence of such thresholds is not clear {\em a priori}. In
contrast to~\cite{Koetter:Vontobel:06:1}, where we discuss cases where we can
guarantee a threshold under LP decoding, here we will show cases where an LP
decoding threshold does not exist. E.g., the sequence of random codes where
each entry in a parity-check matrix is drawn independently from a Bernoulli
distribution with arbitrary nonzero parameter $\theta{}$ does not possess an
SNR-threshold (in the AWGNC case) or an $\varepsilon$-threshold (in the BSC
case). Also, there are ensembles of codes that do not show arbitrary low
probability of decoding error for any SNR (AWGNC) or any positive
$\varepsilon$ (BSC) even if the code rate is allowed to approach zero and the
variable degree is allowed to grow exponentially fast in the block length.

Because LP decoding and MPI decoding are essentially equivalent in the case of
the binary erasure channel (BEC), well-known results about MPI decoding for
the BEC (see e.g.~\cite{Luby:Mitzenmacher:Shokrollahi:Spielman:Stemann:97:1,
Luby:Mitzenmacher:Shokrollahi:98:1}) can be used for making statements about
LP decoding thresholds and therefore we will not deal any further with the BEC
case here.

The paper is structured as follows. After having concluded this introduction
with some remarks about our notation, we will review ML and LP decoding in
Sec.~\ref{sec:ml:and:lp:decoding:1}. The main part of the paper will be
Sec.~\ref{sec:0:neighborhood:based:bounds:1} where we present so-called
$0$-neighborhood-based bounds on the LP decoding threshold. Finally, in
Sec.~\ref{sec:2:neighborhood:based:bounds:1} we will present so-called
$2$-neighborhood-based bounds.

Let us fix some notation. We let $\R$, $\Rp$, and $\Rpp$ be the set of real
numbers, the set of non-negative real numbers, and the set of positive real
numbers, respectively. Moreover, we will use the canonical embedding of the
set $\GF{2} = \{ 0, 1 \}$ into $\R$. The convex hull (see
e.g~\cite{Boyd:Vandenberghe:04:1}) of a set $\set{A} \subseteq \R^n$ is
denoted by $\convhull(\set{A})$. If $\set{A}$ is a subset of $\GF{2}^n$ then
$\convhull(\set{A})$ denotes the convex hull of the set $\set{A}$ after
$\set{A}$ has been canonically embedded in $\R^n$. Similarly, the conic hull
(see e.g~\cite{Boyd:Vandenberghe:04:1}) of a set $\set{A} \subseteq \R^n$ will
be denoted by $\conichull(\set{A})$ and if $\set{A}$ is a subset of $\GF{2}^n$
then $\conichull(\set{A})$ denotes the conic hull of the set $\set{A}$ after
$\set{A}$ has been canonically embedded in $\R^n$. The $i$-th component of a
vector $\vx$ will be called $[\vx]_i$ or $x_i$ and the element in the $j$-th
row and $i$-th column of a matrix $\matr{A}$ will be called
$[\matr{A}]_{j,i}$.

Let $\code{C}$ be a binary linear code defined by a parity-check matrix
$\matr{H}$ of size $m$ by $n$. Based on $\matr{H}$, we define the sets
$\set{I} \defeq \set{I}(\matr{H}) \defeq \{ 1, \ldots, n \}$, $\set{J} \defeq
\set{J}(\matr{H}) \defeq \{ 1, \ldots, m \}$, $\set{I}_j \defeq
\set{I}_j(\matr{H}) \defeq \{ i \in \set{I} \ | \ [\matr{H}]_{j,i} = 1 \}$ for
each $j \in \set{J}$, and $\set{J}_i \defeq \set{J}_i(\matr{H}) \defeq \{ j
\in \set{J} \ | \ [\matr{H}]_{j,i} = 1 \}$ for each $i \in \set{I}$. Moreover,
for each $j \in \set{J}$ we define the codes $\code{C}_j \defeq
\code{C}_j(\matr{H}) \defeq \{ \vx \in \GF{2}^n \ | \ \vect{h}_j \vx^\tr = 0
\text{ (mod $2$)} \}$, where $\vect{h}_j$ is the $j$-th row of
$\matr{H}$. Note that the code $\code{C}_j$ is a code of length $n$ where all
positions not in $\set{I}_j$ are unconstrained. For simplicity of notation, we
will never indicate the parity-check matrix as an argument of $\set{I}$,
$\set{J}$, etc.; it will be clear from the context to what parity-check matrix
we are referring to. Finally, by a family of codes we will mean a sequence of
(deterministicly or randomly generated) codes where the block length goes to
infinity.

\section{ML and LP Decoding}
\label{sec:ml:and:lp:decoding:1}

Let us use the above-mentioned code $\code{C}$ for data transmission over a
binary-input discrete memoryless channel with input alphabet $\set{X} \defeq
\{ 0, 1 \}$, output alphabet $\set{Y}$, and channel law
$P_{Y|X}(y|x)$. Because the channel is memoryless, $P_{\vect{Y} |
\vect{X}}(\vect{y} | \vect{x}) = \prod_{i \in \set{I}} P_{Y | X}(y_i | x_i)$,
where $\vX \defeq (X_1, \ldots, X_n)$, where $\vY \defeq (Y_1, \ldots, Y_n)$,
where the random variable $X_i$ denotes the channel input at time index $i$,
and where the random variable $Y_i$ denotes the channel output at time index
$i$. Upon observing $\vect{Y} = \vect{y}$, the maximum-likelihood (ML)
decoding rule decides for
\begin{align*}
  \hat \vx(\vy)
    &= \arg \max_{\vx \in \code{C}}
         P_{\vect{Y} | \vect{X}}(\vect{y} | \vect{x}). 
\end{align*}
Let the $i$-th log-likelihood ratio $\Gamma_i$, $i \in \set{I}$, be the random
variable
\begin{align*}
  \Gamma_i
    &\defeq
       \Gamma_i(Y_i)
     \defeq
       \log
         \left(
           \frac{P_{Y | X}(Y_i|0)}
                {P_{Y | X}(Y_i | 1)}
         \right)
     \in \R \cup \{ \pm \infty \}
\end{align*}
with realization $\gamma_i \defeq \gamma_i(y_i)$. Then,
noting that
\begin{align*}
  \log P_{Y | X}(y_i | x_i)
    &= -
       \gamma_i x_i
       +
       \log
         P_{Y | X}(y_i | 0),
\end{align*}
ML decoding can also be written as
\begin{align*}
  \hat \vx(\vy)
    &= \arg \min_{\vx \in \code{C}}
         \sum_{i \in \set{I}}
           \gamma_i x_i.
\end{align*}
Because the cost function is linear, and a linear function attains its minimum
at the extremal points of a convex set, this is essentially equivalent to
\begin{align*}
  \hat \vx(\vy)
    &= \arg \min_{\vx \in \convhull(\code{C})}
         \sum_{i \in \set{I}}
           \gamma_i x_i.
\end{align*}
Although this is a linear program, it can usually not be solved efficiently
because its description complexity is usually exponential in the block length
of the code.

However, one might try to solve a relaxation of the above minimization
problem. Noting that $\convhull(\code{C}) \subseteq \bigcap_{j \in \set{J}}
\convhull(\code{C}_j)$ (which follows from the fact that $\code{C} =
\bigcap_{j \in \set{J}} \code{C}_j$), Feldman, Wainwright, and
Karger~\cite{Feldman:03:1, Feldman:Wainwright:Karger:05:1} defined the linear
programming decoder (LP decoder) to be given by the solution of the linear
program
\begin{align}
  \hvomega(\vy)
    &= \arg \min_{\vomega \in \cap_{j \in \set{J}} \convhull(\code{C}_j)}
         \sum_{i \in \set{I}}
           \gamma_i \omega_i.
             \label{eq:lp:decoding:1}
\end{align}
Because of its importance, we will abbreviate the set $\cap_{j \in \set{J}}
\convhull(\code{C}_j)$ by $\fp{P} \defeq \fph{P}{H}$ and call it the
fundamental polytope~\cite{Koetter:Vontobel:03:1, Vontobel:Koetter:05:1:subm}.
The fundamental polytope $\set{P}$ can be expressed using inequalities as
follows~\cite{Feldman:03:1, Feldman:Wainwright:Karger:05:1,
Koetter:Vontobel:03:1, Vontobel:Koetter:05:1:subm}\\[-0.7cm]

{\small
\begin{align*}
  \fp{P}
    &= \left\{
         \vomega \in \R^n
       \ \left| \
         \begin{array}{ll}
           \forall i \in \set{I}:
              0 \leq \omega_i \leq 1 \text{ and } \\
           \forall j \in \set{J},\
           \forall \set{I}'_j \subseteq \set{I}_j,\
           \card{\set{I}'_j} \text{ odd}: \\
              \sum_{i \in \set{I}'_j}
                 \omega_i
               +
               \sum_{i \in (\set{I}_j \setminus \set{I}'_j)}
                 (1-\omega_i)
                 \leq \card{\set{I}_j} - 1
         \end{array}
         \right.
       \right\}.
\end{align*}}%
When analyzing the decoding performance of LP decoding of a binary linear code
that is used for data transmission over a binary-input output-symmetric
channel, we can without loss of generality assume that the all-zeros codeword
was sent. (See also ~\cite{Feldman:03:1}
and~\cite{Feldman:Wainwright:Karger:05:1} that discuss this so-called
``$\code{C}$-symmetry'' property.) We observe that a necessary (but usually
not sufficient) condition that one decides for the all-zeros codeword in
\eqref{eq:lp:decoding:1} is that\footnote{Actually, without changing the
content of the statement in~\eqref{eq:lpd:necessary:cond:1}, we can replace
$\vomega \in \fp{P} \setminus \{ \vect{0} \}$ by $\vomega \in \fp{P}$.}
\begin{align}
  \sum_{i \in \set{I}}
    \gamma_i \omega_i
    &\geq 0
  \quad
  \text{ for all $\vomega \in \fp{P} \setminus \{ \vect{0} \}$}. 
    \label{eq:lpd:necessary:cond:1}
\end{align}
It can easily be seen that this condition is equivalent to the condition that
\begin{align}
  \sum_{i \in \set{I}}
    \gamma_i \omega_i
    &\geq 0
  \quad
  \text{ for all $\vomega \in \conichull(\fp{P}) \setminus \{ \vect{0} \}$}.
    \label{eq:lpd:necessary:cond:2}
\end{align}
The set $\conichull(\fp{P})$, which is the conic hull of the fundamental
polytope, is called the fundamental cone $\fc{K} \defeq \fch{K}{H}$. In terms
of inequalities, $\fc{K}$ can be written as~\cite{Feldman:03:1,
Feldman:Wainwright:Karger:05:1, Koetter:Vontobel:03:1,
Vontobel:Koetter:05:1:subm}
\begin{align*}
  \fp{K}
    &= \left\{
         \vomega \in \R^n
       \ \left| \
         \begin{array}{ll}
           \forall i \in \set{I}:
              0 \leq \omega_i \text{ and } \\
           \forall j \in \set{J},\
           \forall i' \in \set{I}_j:\\
              \omega_{i'}
               -
               \sum_{i \in (\set{I}_j \setminus \{ i' \})}
                 \omega_i
                 \leq 0
         \end{array}
         \right.
       \right\}.
\end{align*}
The condition in~\eqref{eq:lpd:necessary:cond:2} can then be stated as
\begin{align}
  \sum_{i \in \set{I}}
    \gamma_i \omega_i
    &\geq 0
  \quad
  \text{ for all $\vomega \in \fc{K} \setminus \{ \vect{0} \}$}.
    \label{eq:lpd:necessary:cond:3}
\end{align}

We will use the following definition for the block decoding error event under
LP decoding: it is the complement of the event that the all-zeros vector is
the unique solution in~\eqref{eq:lp:decoding:1}.

\section{$0$-Neighborhood-Based Bounds on the Threshold for Regular LDPC Codes}
\label{sec:0:neighborhood:based:bounds:1}

We focus our attention on $(\wcol,\wrow)$-regular LDPC codes, i.e.~codes
defined by parity-check matrices that have uniform column weight $\wcol$ and
uniform row weight $\wrow$. For these type of codes, we present a technique to
obtain bounds on the threshold under LP decoding which we will call
$0$-neighborhood-based bounds; the choice for this name will become clear
later on.

\begin{Assumption}
  In the following, we will always assume that the all-zeros codeword was
  sent, i.e.~we will not explicitly write the conditioning on $\vX = \vect{0}$
  when making statements involving probabilities. \eassumption
\end{Assumption}

Under this assumption, the log-likelihood ratios $\Gamma_1, \ldots, \Gamma_n$
are i.i.d.~random variables and so there is a random variable $\Gamma$ such
that $\Gamma_i \sim \Gamma$ for all $i \in \set{I}$. Moreover, let $\set{G}
\subseteq (\R \cup \{ \pm \infty \})$ be the support of the pdf of $\Gamma$.

\begin{Example}
  \label{ex:gamma:for:different:channels:1}

  Let us discuss the random variable $\Gamma$ for three channels: the
  binary-input additive white Gaussian channel (AWGNC), the binary symmetric
  channel (BSC), and the binary erasure channel (BEC).

  \begin{itemize}
  
    \item AWGNC (with modulation map $0 \mapsto +\sqrt{\Ec}$, $1 \mapsto
    -\sqrt{\Ec}$ and where the added noise has variance $\sigma^2$): $\set{G}
    = \R$ and $\Gamma$ is a continuous random variable that is normally
    distributed with mean $2 \Ec/\sigma^2$ and variance $4 \Ec/\sigma^2$.

    \item BSC (with cross-over probability $\varepsilon$): $\set{G} = \{ \pm G
      \}$ and $\Gamma$ is a discrete random variable that takes on the value
      $G$ with probability $1-\varepsilon$ and the value $-G$ with probability
      $\varepsilon$, where $G \defeq \log\bigl(
      \frac{1-\varepsilon}{\varepsilon} \bigr)$.

    \item BEC (with erasure probability $\epsilon$): $\set{G} = \{ 0, +\infty
      \}$ and $\Gamma$ is a discrete random variable that takes on the value
      $+\infty$ with probability $1-\epsilon$ and the value $0$ with
      probability $\epsilon$.

  \end{itemize}
  \eexample
\end{Example}

\begin{Definition}
  Let
  \begin{align*}
    \Gammapos
      &\defeq 
         + 
         \sum_{i \in \set{I} \atop \Gamma_i \geq 0}
           \Gamma_i
    \quad
    \text{ and }
    \quad
    \Gammaneg
       \defeq
         -
         \sum_{i \in \set{I} \atop \Gamma_i < 0}
           \Gamma_i
  \end{align*}
  be random variables with realizations $\gammapos$ and $\gammaneg$,
  respectively. Note that $\Gammapos \geq 0$ and $\Gammaneg \geq 0$ w.p.~$1$.
  \edefinition
\end{Definition}

\begin{Lemma}
  \label{lemma:Gammapos:Gammaneg:lln:1}

  With probability one we have
  \begin{alignat*}{2}
    &
    \lim_{n \to \infty}
      \frac{\Gammapos}{n}
        &= +
           \Expec[\Gamma \, | \, \Gamma {\geq} 0]
           \cdot
           \Prob(\Gamma {\geq} 0)
        &= +
           \int_{0^{-}}^{+\infty}
             \!\!\!\!\!\!
             \gamma \,
             p_{\Gamma}(\gamma)
           \dint{\gamma}, \\
    &
    \lim_{n \to \infty}
      \frac{\Gammaneg}{n}
        &= -
           \Expec[\Gamma \, | \, \Gamma {<} 0]
           \cdot
           \Prob(\Gamma {<} 0)
        &= -
           \int_{-\infty}^{0^{-}}
             \!\!\!
             \gamma \,
             p_{\Gamma}(\gamma)
           \dint{\gamma}.
  \end{alignat*}
\end{Lemma}

\begin{Proof}
  Follows easily from the weak law of large numbers.
\end{Proof}

\begin{Lemma}
  \label{lemma:Gammapos:Gammaneg:ratio:bound:1}

  Consider a code with a $(\wcol, \wrow)$-regular parity-check matrix. Let
  $\vgamma \in \set{G}^n$. A necessary condition that the LP decoder decides
  in favor of the all-zeros codeword is
  \begin{align*}
    \frac{\gammapos}{\gammaneg}
      &\geq
         \wrow - 1.
  \end{align*}
\end{Lemma}

\mbox{}

\begin{Proof}
  We saw in~\eqref{eq:lpd:necessary:cond:3} that a necessary condition for LP
  decoding to decide in favor of the all-zeros codeword is that $\sum_{i \in
  \set{I}} \gamma_i \omega_i \geq 0$ for all $\vomega \in \fc{K} \setminus \{
  \vect{0} \}$. Let us construct a vector $\vomega \in \R^n$ as follows:
  \begin{align}
    \omega_i
      &\defeq 
         \begin{cases}
           \frac{1}{\wrow-1} & \text{if $\gamma_i \geq 0$} \\
           1                 & \text{if $\gamma_i < 0$}
         \end{cases}.
           \label{eq:zero:neighborhood:decoding:1}
  \end{align}
  It can easily be seen that $\vomega \in \fch{K}{H}$.\footnote{The vector
  $\vomega$ can be seen as a generalization of the so-called canonical
  completion~\cite{Koetter:Vontobel:03:1, Vontobel:Koetter:05:1:subm}, however
  instead of assigning values according to the graph distance with respect to
  a single node, we assign values according to the graph distance with respect
  to the set of nodes where $\gamma_i$ is negative. Moreover, we assign only
  the values $1$ and $1 / (\wrow - 1)$.} We obtain the necessary condition
  \begin{align*}
    0 
      &\leq
         \sum_{i \in \set{I}}
           \gamma_i \omega_i
       = - 1 \cdot \gammaneg
         + \frac{1}{\wrow-1} \cdot \gammapos,
  \end{align*}
  which is equivalent to the necessary condition in the lemma statement.
\end{Proof}

\begin{Theorem}
  \label{th:Gammapos:Gammaneg:ratio:bound:lln:1}

  Consider a family of codes that have $(\wcol, \wrow)$-regular parity-check
  matrices. In the limit $n \to \infty$, a necessary condition such that the
  LP decoder decides in favor of the all-zeros codeword with probability one
  is
  \begin{align*}
    -
    \frac{\Expec[\Gamma \, | \, \Gamma {\geq} 0]}
         {\Expec[\Gamma \, | \, \Gamma {<} 0]}
    \cdot
    \frac{\Prob(\Gamma {\geq} 0)}
         {\Prob(\Gamma {<} 0)}
      &\geq
         \wrow - 1,
  \end{align*}
  or, equivalently,
  \begin{align*}
    -
    \frac{\int_{0^{-}}^{+\infty}
            \gamma \,
            p_{\Gamma}(\gamma)
          \dint{\gamma}}
         {\int_{-\infty}^{0^{-}}
            \gamma \,
            p_{\Gamma}(\gamma)
          \dint{\gamma}}
      &\geq
         \wrow - 1.
  \end{align*}
\end{Theorem}

\mbox{}

\begin{Proof}
  This follows upon combining Lemmas~\ref{lemma:Gammapos:Gammaneg:lln:1}
  and~\ref{lemma:Gammapos:Gammaneg:ratio:bound:1}.
\end{Proof}

\begin{Corollary}
  \label{cor:bsc:threshold:ub:1}

  Consider the setup of Th.~\ref{th:Gammapos:Gammaneg:ratio:bound:lln:1}.  If
  the memoryless channel is a BSC with cross-over probability $\varepsilon$
  then the necessary condition in
  Th.~\ref{th:Gammapos:Gammaneg:ratio:bound:lln:1} reads
  \begin{align*}
    \varepsilon
      &\leq \frac{1}{\wrow}.
  \end{align*}
\end{Corollary}

\mbox{}

\begin{Proof}
  For a BSC with cross-over probability $\varepsilon$ we obtain
  \begin{align*}
    \Expec[\Gamma \, | \, \Gamma {\geq} 0] \cdot \Prob(\Gamma {\geq} 0)
      &= + G \cdot (1-\varepsilon), \\
    -\Expec[\Gamma \, | \, {\Gamma} {<} 0] \cdot \Prob(\Gamma {<} 0) 
      &= - (-G) \cdot \varepsilon = + G \cdot \varepsilon
  \end{align*}
  where $G$ is defined as in Ex.~\ref{ex:gamma:for:different:channels:1}.
  Therefore, the condition in Th.~\ref{th:Gammapos:Gammaneg:ratio:bound:lln:1}
  is that $\frac{G (1-\varepsilon)}{G \varepsilon} \geq \wrow - 1$,
  i.e.~$\varepsilon \leq \frac{1}{\wrow}$.
\end{Proof}

\begin{figure}
  \begin{center}
    \epsfig{file=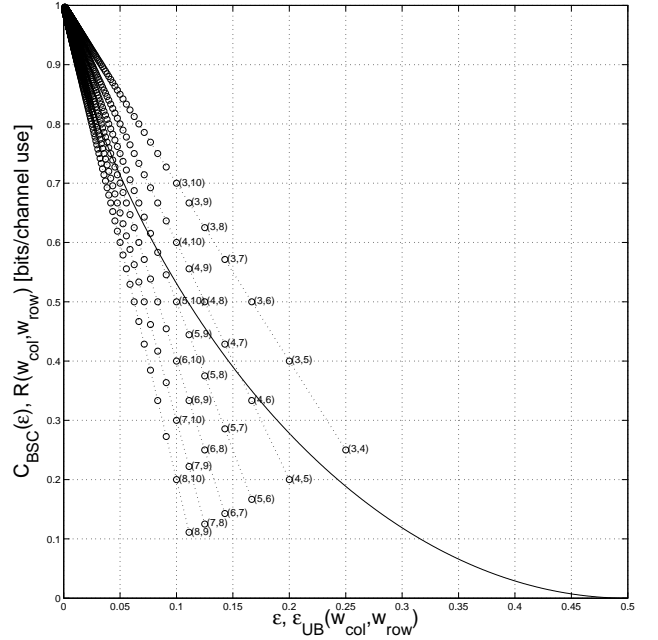, width=0.95\linewidth}

    \caption{The solid line shows the capacity $C_{\mathrm{BSC}}(\varepsilon)$
      of a BSC as a function of the cross-over probability $\varepsilon$. The
      circles have the following meaning: the circle with label $(\wcol,
      \wrow)$ shows the point $\bigl( R(\wcol, \wrow),
      \varepsilon_{\mathrm{UB}}(\wcol, \wrow) \bigr)$, where $R(\wcol, \wrow)$
      and $\varepsilon_{\mathrm{UB}}(\wcol, \wrow)$ are the designed rate and
      the threshold upper bound from Cor.~\ref{cor:bsc:threshold:ub:1},
      respectively, for $(\wcol,\wrow)$-regular LDPC codes.}

    \label{fig:threshold:lp:reg:codes:1:1}
  \end{center}
\end{figure}

\begin{Example}
  Fig.~\ref{fig:threshold:lp:reg:codes:1:1} tries to capture some of the
  implications of Th.~\ref{th:Gammapos:Gammaneg:ratio:bound:lln:1} /
  Cor.~\ref{cor:bsc:threshold:ub:1}. The circles in the plot that are to the
  left of the capacity curve yield non-trivial upper bounds on the error
  correction capability of LP decoding. (Note that the designed rate of a
  $(\wcol,\wrow)$-regular LDPC code is $1 - \frac{\wcol}{\wrow}$ and that the
  actual rate is lower bounded by this quantity.) \eexample
\end{Example}

\begin{figure}
  \begin{center}
    \epsfig{file=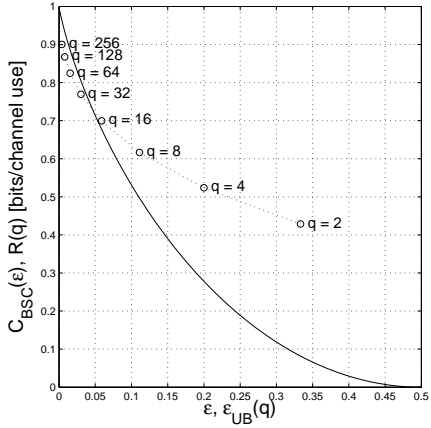, width=0.65\linewidth}

    \caption{The solid line shows the capacity $C_{\mathrm{BSC}}(\varepsilon)$
      of a BSC as a function of the cross-over probability $\varepsilon$. The
      circles have the following meaning: the circle with label $q$ shows the
      point $\bigl( R(q), \varepsilon_{\mathrm{UB}}(q) \bigr)$, where $R(q)$
      is the rate of the $\mathrm{PG}(2,q)$-based code and where
      $\varepsilon_{\mathrm{UB}}(q)$ is the threshold upper bound from
      Cor.~\ref{cor:bsc:threshold:ub:1} for a family of
      $(\wcol(q),\wrow(q))$-regular LDPC codes. (See also the main text for
      further explanations.)}

    \label{fig:threshold:lp:pg:codes:1:1}
  \end{center}
\end{figure}

\begin{Example}
  \label{ex:threshold:all:equal:weight:rows:1}
  
  It is a surprising fact that the bounds in
  Th.~\ref{th:Gammapos:Gammaneg:ratio:bound:lln:1} and
  Cor.~\ref{cor:bsc:threshold:ub:1} do not depend on the variable degree
  $\wcol$ at all. In particular, this implies that decoding would fail even if
  we choose extremely large variable degrees. For example we might consider a
  sequence of codes defined by parity-check matrices that contain {\em all}
  rows of a given weight $\wrow$. Clearly, a specific code $\code{C}_n$ of
  this sequence is a $(\intextbinomial{n-1}{\wrow-1},\wrow)$-regular code
  which contains either one ($\code{C}_n = \{ \vect{0} \}$) or two codewords
  ($\code{C}_n = \{ \vect{0}, \vect{1} \}$) depending on $k$ being odd or
  even. Thus, while the rate of this code sequence approaches zero, LP decoding
  will not succeed for $\varepsilon > \frac{1}{\wrow}$. \eexample
\end{Example}

\begin{Example}
  \label{ex:threshold:unbounded:row:weight:1}

  Th.~\ref{th:Gammapos:Gammaneg:ratio:bound:lln:1} and
  Cor.~\ref{cor:bsc:threshold:ub:1} can easily be extended to families of
  codes where the row weight grows as a function of $n$; let us call theses
  extensions Th.~\ref{th:Gammapos:Gammaneg:ratio:bound:lln:1}' and
  Cor.~\ref{cor:bsc:threshold:ub:1}'. It is clear from
  Cor.~\ref{cor:bsc:threshold:ub:1}' that there cannot be an LP decoding
  threshold for the BSC if the row weight grows unboundedly. Moreover, coming
  back to the code family in Ex.~\ref{ex:threshold:all:equal:weight:rows:1},
  if $\wrow$ is allowed to grow with $n$, LP decoding will fail as $n \to
  \infty$ despite the variable degree being exponentially larger than the
  check degree.
\end{Example}

\begin{Example}
  Similarly, the family of random codes where entries in a parity-check matrix
  are drawn independently from a Bernoulli($\theta$) distribution will not
  only have poor threshold performance under LP decoding but will fail with
  high probability as the code length approaches infinity for {\em{any}}
  symmetric channel for which the expression
  \begin{align*}
    -
    \frac{\Expec[\Gamma \, | \, \Gamma {\geq} 0]}
         {\Expec[\Gamma \, | \, \Gamma {<} 0]}
    \cdot
    \frac{\Prob(\Gamma {\geq} 0)}
         {\Prob(\Gamma {<} 0)}
  \end{align*}
  is (upper) bounded. The result follows from the observation that the weight
  of the rows in $\matr{H}$ is exponentially concentrated around $\theta n$.
  Indeed, given a vector of log-likelihood ratios, the vector with components
  $\frac{1}{n\theta - \delta - 1}$ in positions where $\gamma_i$ is
  non-negative and $1$ in the remaining positions is inside $\fch{K}{H}$ with
  high probability for $\delta > 0$ and $n \to \infty$. \eexample
\end{Example}

While the above considerations give some insight in the asymptotic behavior of
of decoding error for LP decoding, the characterization and spirit of
Th.~\ref{th:Gammapos:Gammaneg:ratio:bound:lln:1} is essentially
combinatorial.

\begin{Example}
  We saw in Ex.~\ref{ex:threshold:unbounded:row:weight:1} that for any family
  of codes where the row weight grows as a function of $n$,
  Cor.~\ref{cor:bsc:threshold:ub:1}' implies that there cannot be an LP
  decoding threshold for the BSC. A special case, though, arises when the rate
  of the code family under consideration goes to $1$ when $n \to \infty$
  because then also the best code family under the best possible decoding
  algorithm can only correct a vanishing fraction of bit flips as $n \to
  \infty$. A family were the rate goes to $1$ as $n \to \infty$ is the family
  of type-I $\mathrm{PG}(2,q)$-based codes, cf.~\cite{Kou:Lin:Fossorier:01:1};
  in the context of LP decoding, these codes were analyzed
  in~\cite{Vontobel:Smarandache:Kiyavash:Teutsch:Vukobratovic:05:1,
  Vontobel:Smarandache:05:1}. A code from this family is indexed by $q$ (where
  $q \defeq 2^s$ for some positive integer $s$), has length $n(q) \defeq
  q^2+q+1$, rate $R(q) \defeq 1 - (3^s + 1)/(q^2+q+1)$, and $\wcol(q) =
  \wrow(q) = q+1$.

  Fig.~\ref{fig:threshold:lp:pg:codes:1:1} shows the following: for each $q$
  we plot the point $(R(q), \varepsilon_{\mathrm{UB}}(q))$, where $R(q)$ is
  the rate of the $\mathrm{PG}(2,q)$-based code and where
  $\varepsilon_{\mathrm{UB}}(q)$ is the LP decoding threshold upper bound
  from Cor.~\ref{cor:bsc:threshold:ub:1} for a $(\wcol(q), \wrow(q))$-regular
  family of codes. Care must be taken when giving an interpretation to this
  figure since the $\mathrm{PG}(2,q)$-based codes are finite-length codes for
  finite $q$.
  \eexample
\end{Example}

We leave it as an exercise for the reader to generalize the results in this
section to irregular LDPC codes.

\section{$2$-Neighborhood-Based Bounds on the Threshold for Regular LDPC Codes}
\label{sec:2:neighborhood:based:bounds:1}

\begin{figure}
  \begin{center}
    \epsfig{file=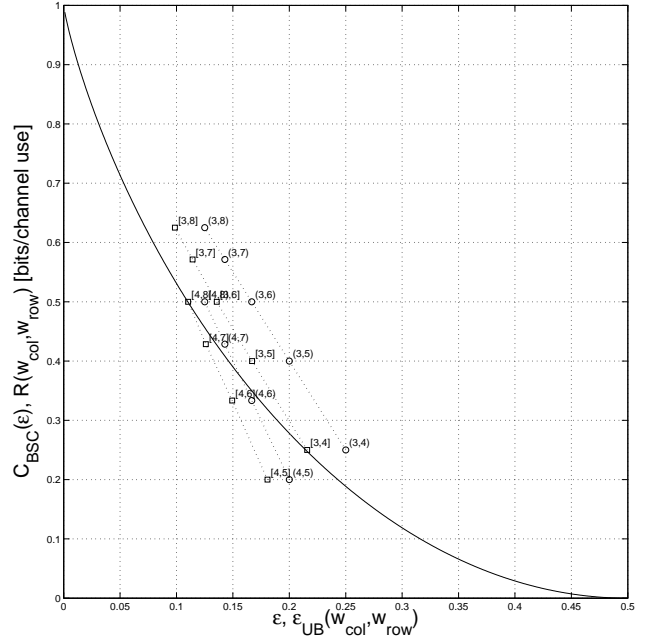, 
            width=0.95\linewidth}

    \caption{The solid line shows the capacity $C_{\mathrm{BSC}}(\varepsilon)$
      of a BSC as a function of the cross-over probability $\varepsilon$. The
      circles have the same meaning as in
      Fig.~\ref{fig:threshold:lp:reg:codes:1:1}. The squares have the
      following meaning: the square with label $q$ shows the point $\bigl(
      R(q), \varepsilon_{\mathrm{UB}}(q) \bigr)$, where $R(q)$ is the rate of
      the $\mathrm{PG}(2,q)$-based code and where
      $\varepsilon_{\mathrm{UB}}(q)$ is the $2$-neighborhood-based threshold
      upper bound from Sec.~\ref{sec:2:neighborhood:based:bounds:1} for a
      family of $(\wcol(q),\wrow(q))$-regular LDPC codes. (See also the main
      text for further explanations.)}

    \label{fig:threshold:lp:reg:codes:neighborhood:2:1:1}
  \end{center}
\end{figure}

Because the assignment of a value to $\omega_i$
in~\eqref{eq:zero:neighborhood:decoding:1} was only based on the value of
$\gamma_i$, we call the resulting bound in Cor.~\ref{cor:bsc:threshold:ub:1} a
$0$-neighborhood-based bound. (Of course, the way we assigned a value to every
$\omega_i$ in~\eqref{eq:zero:neighborhood:decoding:1} can also be seen as a
very simplistic, and usually sub-optimal way, of solving the linear program
in~\eqref{eq:lp:decoding:1}.) It is natural to try to formulate more
sophisticated assignments of a value to $\omega_i$. The next simplest approach
is to formulate a rule that does not depend on $\gamma_i$ only, but also on
$\gamma_{i'}$ where $i'$ ranges over all variable nodes at Tanner graph
distance $2$ from variable node $i$. The resulting bounds on the threshold
will therefore be called $2$-neighborhood-based bounds.

\begin{figure}
  \begin{center}
    \epsfig{file=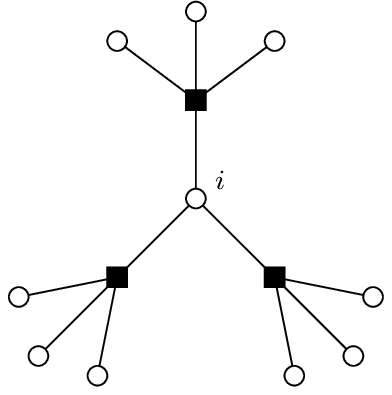, width=0.30\linewidth}
    \quad\quad
    \epsfig{file=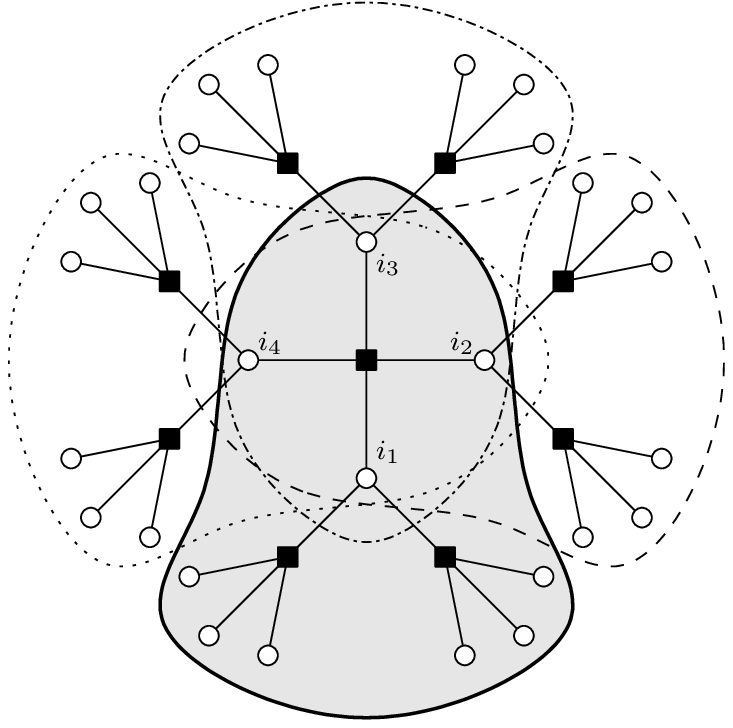, width=0.50\linewidth}

    \caption{Left: $2$-neighborhood. Right: overlapping
      $2$-neighborhoods. (See Sec.~\ref{sec:2:neighborhood:based:bounds:1} for
      more explanations.)}

    \label{fig:tg:two:neighborhood:1}
  \end{center}
\end{figure}

For $i \in \set{I}$ let $\set{N}^{(2)}_i$ be the subset of $\set{I}$ that
includes all variable nodes $i'$ with Tanner graph distance at most $2$ from
$i$. In the following, we assume that the Tanner graph under consideration has
girth at least six. In the case of a $(\wcol, \wrow)$-regular LDPC codes, this
implies that the set $\set{N}^{(2)}_i$ has size $\card{\set{N}^{(2)}_i} = 1 +
\wcol (\wrow-1)$. (Fig.~\ref{fig:tg:two:neighborhood:1} (left) shows part of a
$(3,4)$-regular code, i.e.~node $i$, all check nodes at Tanner graph distance
$1$ from $i$, and all variable nodes at Tanner graph distance $2$ from $i$.)

\begin{Definition}
  \label{def:2:neighborhood:decoding:rule:1}

  Let $\vGamma'_i$ be the vector that contains all the random variables $\{
  \Gamma_{i'} \}_{i' \in \set{N}^{(2)}_i}$ and let $\vgamma'_i$ be its
  realization. Our new rule (that
  replaces~\eqref{eq:zero:neighborhood:decoding:1}) for defining a vector
  $\vomega \in \R^n$ is now $\omega_i \defeq \alpha_{\vgamma'_i}$, where
  $\alpha_{\vgamma'_i} \in \Rp$ for all $\vgamma'_i \in
  \set{G}^{\card{\set{N}^{(2)}_i}}$ is chosen such that for all possible
  $\vgamma \in \set{G}^n$ we obtain a vector $\vomega$ that lies in the
  fundamental cone.
\end{Definition}

\begin{Lemma}
  Assume that we have such values $\{ \alpha_{\vgamma'_i} \}_{\vgamma'_i}$ as
  defined in Def.~\ref{def:2:neighborhood:decoding:rule:1}. With probability
  one with have
  \begin{align*}
    \lim_{n \to \infty}
      \frac{1}{n}
      \sum_{i \in \set{I}}
        \gamma_i \omega_i
      &= \sum_{\vgamma'_i}
           p_{\vgamma'_i}        
           \alpha_{\vgamma'_i},
  \end{align*}
  where
  \begin{align*}
    p_{\vgamma'_i}
      &\defeq
         \Pr\big(\{ \Gamma_{i'} \}_{i' \in \set{N}^{(2)}_i}
             = \{ \gamma_{i'} \}_{i' \in \set{N}^{(2)}_i}\big)
       = \prod_{i' \in \set{N}^{(2)}_i}
           \Pr(\Gamma = \gamma_{i'}).
  \end{align*}
\end{Lemma}

\begin{Proof}
  Follows from the fact that $\gamma_i \omega_i$ depends only on finitely many
  $\gamma_{i'}$ from $\{ \gamma_{i'} \}_{i' \in \set{I}}$ and from the use of
  the weak law of large numbers.
\end{Proof}

Consider a BSC with cross-over probability $\varepsilon$. An upper bound on
the LP decoding threshold for the BSC is then given by the infimum of all
$\varepsilon$ were we are able to find an assignment in
Def.~\ref{def:2:neighborhood:decoding:rule:1} such that $\lim_{n \to \infty}
\frac{1}{n} \sum_{i \in \set{I}} \gamma_i \omega_i$ is negative with
probability one. Finding such assignments can e.g.~be done by solving a linear
program that roughly looks as follows
\begin{alignat*}{2}
  &
  \text{min.}\quad
    &&\sum_{\vgamma'_i}
        p_{\vgamma'_i}        
        \alpha_{\vgamma'_i} \\
  &
  \text{subj.~to}\quad
    &&\text{$\alpha_{\tilde \vgamma'_i} = 1$, } \\
    &&&\text{and for each $\vgamma$ the assignment always results in a} \\
    &&&\text{non-zero vector that lies in the fundamental cone},
\end{alignat*}
where $\tilde \vgamma'_i$ is an arbitrary assignment of values to
$\vgamma'_i$, e.g.~$\gamma_{i'} = -G$ for all $i' \in \set{N}^{(2)}_i$, where
$G$ is defined as in Ex.~\ref{ex:gamma:for:different:channels:1}.

The Tanner graph in Fig.~\ref{fig:tg:two:neighborhood:1} (left) has many
symmetries that can be used to simplify the above linear program. E.g.~if
there is a graph isomorphism that maps an assignments
${\overline{\vgamma}}'_i$ to an assignment
${\overline{\overline{\vgamma}}}'_i$, then without loss of generality we can
assume that $\alpha_{{\overline{\vgamma}}'_i} =
\alpha_{{\overline{\overline{\vgamma}}}'_i}$. In this way, the dimensionality
of the above linear program can be reduced significantly. Without going into
the details, the requirement that the resulting assignment always results in a
vector in the fundamental cone can be simplified by introducing some auxiliary
variables according to overlapping $2$-neighborhoods as is sketched in
Fig.~\ref{fig:tg:two:neighborhood:1} (right).

\begin{Example}
  Fig.~\ref{fig:threshold:lp:reg:codes:neighborhood:2:1:1} shows the improved
  upper bounds for $(\wcol,\wrow)$-regular code families.
  \eexample
\end{Example}

Whereas the above approach results in relatively small linear programs for
small $\wcol$ and $\wrow$, similar $4$-, $6$-, $8$-, etc., neighborhood-based
approaches seem to be computationally much more demanding. We leave it as an
open question to see if there are ways to handle also these cases in an
efficient numerical way.

\section*{Acknowledgments}

P.O.V.'s research was supported by NSF Grant CCF-0514801. R.K.'s research was
supported by NSF Grant CCF-0514869.

\end{document}